%% file: socialcom2010.tex
\documentclass[conference]{IEEEtran}

\usepackage[numbers,sort]{natbib}
\usepackage[pdftex]{graphicx}
\usepackage[cmex10]{amsmath}
\usepackage{multirow}
\usepackage{algorithmic}
\usepackage{algorithm}
\usepackage[english]{babel}
\usepackage{amssymb}
\usepackage{colortbl}
\usepackage[table]{xcolor}

\newcommand{\VSP}[0]{\vspace{-0.43cm}}

\begin{document}

\title{Precursors and Laggards: An Analysis of Semantic Temporal Relationships on a Blog Network}

\author{\IEEEauthorblockN{Telmo Menezes}
\IEEEauthorblockA{CREA \& ISCPIF\\
CNRS\\
ISC - 57-59, rue Lhomond\\
F-75005 Paris, France\\
Email: telmo@telmomenezes.com}
\and
\IEEEauthorblockN{Camille Roth}
\IEEEauthorblockA{CAMS \& ISCPIF\\
CNRS-EHESS\\
54, bd Raspail\\
F-75006 Paris, France\\
Email: roth@ehess.fr}
\and
\IEEEauthorblockN{Jean-Philippe Cointet}
\IEEEauthorblockA{INRA-SenS \& ISCPIF\\
INRA-SenS, Universit\' e Paris-Est, Bois de l'Etang\\
5, Bd Descartes, Champs sur Marne\\
F-77454 Marne-la-Vall\' ee, France\\
Email: jean-philippe.cointet@polytechnique.edu}}

\maketitle

\begin{abstract}

We explore the hypothesis that it is possible to obtain information about the dynamics of a blog network by analysing the temporal relationships between blogs at a semantic level, and that this type of analysis adds to the knowledge that can be extracted by studying the network only at the structural level of URL links. We present an algorithm to automatically detect fine-grained discussion topics, characterized by n-grams and time intervals. We then propose a probabilistic model to estimate the temporal relationships that blogs have with one another. We define the precursor score of blog A in relation to blog B as the probability that A enters a new topic before B, discounting the effect created by asymmetric posting rates. Network-level metrics of precursor and laggard behavior are derived from these dyadic precursor score estimations. This model is used to analyze a network of French political blogs. The scores are compared to traditional link degree metrics. We obtain insights into the dynamics of topic participation on this network, as well as the relationship between precursor/laggard and linking behaviors. We validate and analyze results with the help of an expert on the French blogosphere. Finally, we propose possible applications to the improvement of search engine ranking algorithms.
\end{abstract}

\IEEEpeerreviewmaketitle

\input{soccom10-introduction}

% An example of a double column floating figure using two subfigures.
% (The subfig.sty package must be loaded for this to work.)
% The subfigure \label commands are set within each subfloat command, the
% \label for the overall figure must come after \caption.
% \hfil must be used as a separator to get equal spacing.
% The subfigure.sty package works much the same way, except \subfigure is
% used instead of \subfloat.
%
%\begin{figure*}[!t]
%\centerline{\subfloat[Case I]\includegraphics[width=2.5in]{subfigcase1}%
%\hfil
%\subfloat[Case II]{\includegraphics[width=2.5in]{subfigcase2}%
%\label{fig_second_case}}}
%\caption{Simulation results}
%\label{fig_sim}
%\end{figure*}
%
% Note that often IEEE papers with subfigures do not employ subfigure
% captions (using the optional argument to \subfloat), but instead will
% reference/describe all of them (a), (b), etc., within the main caption.

\input{soccom10-topics}

\input{soccom10-probabilistic}

\input{soccom10-discussion}

\input{soccom10-conclusion}

%\mathcal{G}
\section*{Acknowledgments}
This work has been
partially supported by the French National Agency of Research (ANR)
through grant {\nobreak{``Webfluence''}} \#ANR-08-SYSC-009. We thank Cl\'emence Lerondeau and Guilhem Fouetillou from {\sc Linkfluence} for their help in qualitatively assessing the data.


\end{document}

%% file: soccom10-introduction.tex
%\documentclass[10pt]{amsart}
%\usepackage{geometry}                % See geometry.pdf to learn the layout options. There are lots.
%\geometry{a4paper}                   % ... or a4paper or a5paper or ... 
%%\geometry{landscape}                % Activate for for rotated page geometry
%%\usepackage[parfill]{parskip}    % Activate to begin paragraphs with an empty line rather than an indent
%\usepackage{graphicx}
%\usepackage{amssymb}
%\usepackage{natbib}
%\usepackage{epstopdf}
%\DeclareGraphicsRule{.tif}{png}{.png}{`convert #1 `dirname #1`/`basename #1 .tif`.png}

%\title{Leaders and followers}
%\author{SocialCom 2010}
%%\date{}                                           % Activate to display a given date or no date

%\begin{document}
%\maketitle
\section{Introduction}

%is ubiquitous and rather obvious in most social systems. 
For cultural anthropologists, understanding fads, trends, or, generally, cultural similarity, essentially comes to explaining ``the capacity of some representations to propagate until becoming precisely cultural, that is, revealing the reasons of their contagiosity'' \citep{lenc:cult}.
This type of research programme admittedly assumes the possibility of, on one hand, describing representations in a consistent manner, and, on the other hand, apprehending processes of social mediation. 
Defining consistent cultural items is indeed crucial to describe adoption of similar ideas, behaviors, opinions, topics, etc. --- the literature proposes here a large variety of concepts, such as using same bags of terms, having identical opinion vectors, duplicating references (for instance to digital content such as online video or news articles, tagged by the same URL) or, more loosely, being ``infected'' by spreading ``memes''.
Second, describing social mediation requires to understand jointly how some types of social network configurations and some types of interactions may or may not favor the transmission, reproduction or adoption of behaviors, ideas, etc. Again, a vast amount of research has been concerned with normative models or descriptive protocols aimed at understanding which kind of individuals were more or less likely to pass on some pieces of information, and which type of network positions could favor the diffusion of some items.

By relying on large-scale datasets on which individuals talk about what and when, specifically in online communities, social computing has recently contributed to this broad research programme by intensively developing two pragmatic streams of study: detection of ``topics'', and characterization of ``informational cascades''.
Studies focused on topic detection explore bursts and regularities of behavior or term use \citep[e.g.,][]{klei:burs}, sometimes in order to infer trends in the general population \citep{gins:dete,asur:pred}. 
In all these studies, cultural representations are assumed to be extremely atomic, \hbox{i.e.} based on a single behavior (a vote), item (a reference, a URL), apprehending cultural contagion pretty much similarly to disease contagion --- to the notable exception of \citep{lesk:meme} who gather similar sentences into clusters of quotes, getting closer to the polymorphism of cultural representations emphasized by anthropologists.

On the other hand, studies on informational cascades currently adopt a structural stance, migrating from the ``two-step-model'' to more recent arguments underlining the importance of more horizontal, less hierarchical patterns \citep{watt:infl,Cha:2010p2742}. Importantly, in this persective, information flows and diffusion paths are characterized along a given social network, available \emph{a priori}. In many cases however, and certainly in blogs in particular, much of the information regarding the whole underlying interaction infrastructure is simply missing (be it in terms of news media readership, email exchanges and broadly any type of non-blog-based online conversation, phone calls, etc.).

In this paper, we aim at bridging these rather separate streams by adopting (i) a looser view on representations, as stories or cultural attractors \citep{sper:expl,sper:mode} rather than atomic items and, (ii) by considering information sources, in our case bloggers, as sensors in a social system -- in particular as representatives of topics discussed in the society -- so as to suggest possible/implicit information diffusion flows or, at least, precedence relationships. 
As an aside, the current contribution also considers \emph{observed} social networks as effects rather than just causes of information diffusion.

We thus propose to identify topic classes, exhibit temporal precedence relations between sources based on \emph{significant plausibility} for an individual to address a topic before others do, and eventually compare this structure with the partial network of interactions constituted by explicit links among bloggers.  Classical authority measures are found to have only a weak correlation with our approach, which rather exhibits potential online whistleblowers. The next section presents an overview of the relevant literature, while Sec.~\ref{sec3} details the empirical protocol used to identify topics. Sec.~\ref{sec4} then describes our approach to compute probable precedence relationships; results are discussed and reframed in Sec.~\ref{sec5}.

\section{Related work}\label{sec:related}
\subsection{Temporal detection of topics/bursts.}
Topic characterization from (online) text corpora generally relies on \emph{terms}, \emph{n-grams} (\hbox{i.e.} a basic linguistic unit of $n$ terms) or sentence segments.  Once basic text units have been defined and extracted, topics are appraised both quantitatively and temporally, essentially by describing ``how much on which period of time they are being discussed''.  This led to distinguishing bursts of interest (``spikes'') \cite{klei:burs}, as opposed to continuous discussions  (``chatters'') around topics \citep{gruh:info}. Models of the temporal \citep{Balog:2006p2268} or spatial \citep{lloy:news} regularities in the usage of topics have been subsequently developed, up to infering and predicting accurate information regarding the whole population behavior \citep{gins:dete,asur:pred}.

Another stream of research has focused on improving the qualification of topics: for instance, by detecting whether issues are addressed in a positive light or not \citep[the so-called field of ``sentiment analysis'', see][among others]{Mishne:2006p2282}; or, closer to our issues, by managing to group portions of text into classes of similar content \citep{lesk:meme} --- thereby implicitly addressing one common critique among social scientists regarding the atomism of ``memes'' as cultural items.

\subsection{Precedence and influence}
%Empirical studies on influence data obviously relies on precedence relationships first, often following the assumption that an individual may only be said to be \emph{influenced} if she does something after another individual did. As such, influence is generally appraised from the perspective of interaction networks, using relational information to characterize contagion paths. 
Empirical studies of influence generally rely on interaction networks, using relational information to characterize contagion paths, and following a long tradition in mathematical sociology of social network-based models of information diffusion. As regards blogspace in particular, after initial descriptions of the underlying social network structure \citep[e.g.][who also discuss bursty behavior in link creation]{kuma:burs}, \citep{Leskovec:sdm} has been one of the first studies to specifically focus on the structure of link cascades. In a previous study, \cite{coin:soci} describe more precisely local influence patterns such as the relationship between  \hbox{e.g.} holistic patterns and the weakness of links, in Granovetter's sense. 
\citep{Java:2006p1951}, on the other hand, use various social network structures to show that possible influence of a given blog is best described by strictly structural page-rank-style measures.

%Same issue: see also methods exclusively relying on more or less elaborate constructions upon the different possible networks of inter-blog links: in 

Since influence is obviously related to precedence relationships, % (whereby influence could only be invoked when an individual does something after another individual or group of individuals), 
several papers focus rather on temporal behavioral precedence.  For instance, the authors of \citep{koss:stru} exhibit explicit temporal dependencies on a email transmission network by characterizing possible shortcuts in information paths, because a dyad (A,B) could communicate less quickly than (A,C) and (C,B) separately do. 

In terms of intertwining social network structure and precedence/influence, the relationship between topology and precursors or laggards had also been explored in \citep{vale:soci}, but with the assumption that the social network is known a priori, and by monitoring the adoption of a unique yes-or-no behavior.
As said before, it is likely that a lot of information about the social structure is missing in most of the above studies, which consider the (given) social network as the substrate of information propagation. By assuming that the social structure describes only a non-significant fraction of all possible interaction links and contagion paths in the context of (for instance) political discussions, we basically wish to suggest that, here, the social network could just be a secondary material in the study of contagion. 

Some studies do exactly so and exhibit influence relationships from usage information only: for instance in \citep{zhou:topi} a Markov Chain Model is used to characterize which topics are most likely to transition into others, using data extracted from scientific bibliographic databases. Back to blogs, ``probable'' content diffusion paths could be exhibited in \citep{adar:impl} by using classifiers based upon blog features: for instance, having similar citing and content posting patterns; however, the analysis does not seem to make use of topic dynamics \emph{per se}. 
Another reference \citep{Java:2006p1948} introduces an analysis which integrates more semantics, essentially in order to design automatic feed recommenders --- which appears nonetheless to be still based on structural features (in-degree statistics) even if a filter is applied over general topics (politics vs. IT, etc.).

On the whole, and in the context of partial social network information, the issue of the detection of implicit, non-structural influence flows using \emph{temporal} precedence in addressing topics remains a pending question.

%{\small \subsection{Rubbish}
%Because of its essentially temporal nature, topic detection is influenced by  content production dynamics in a given online community.  In terms of blogspace, some earlier studies focused on the periodicity of blog posting, or exhibited citation behavior (decreasing probability of being cited along time) \citep{Leskovec:sdm}. 
%``On the bursty evolution of blogspace'' extends this dynamics analysis to the social network itself, exhibiting bursts in the creation of links \citep{kuma:burs}. 
%}

%% file: soccom10-topics.tex
%\section{Topic Detection}
\section{Unit of activity detection}
\label{sec3}
%We are interested in identifying topics of discussion for which we can later analyse the temporal relationships of their participants. Such topics must have two characteristics to be relevant to our analysis: to have well defined time boundaries within our observation period and to be maintained by the participation of several blogs. 

%
%minimum of blogs participating
%must be bursty (with a minimum intensity, but this is an empirical constraint)
%25\% of the timeline
%direct experimentations.

We are interested in identifying topics of discussion for which we can later analyse the temporal relationships of their participants. Such topics must have two characteristics to be relevant to our analysis: to have well defined time boundaries within our observation period and to be maintained by the participation of several blogs. If these two constrained are respected then we are observing what we will call a well defined ``unit  of activity''. We empirically define a method that identify bursty topics which meet these  constraints.

In~\cite{lesk:meme}, research related to the problem of topic detection is classified into two main categories: probabilistic models to identify long-range trends in general topics and the use of rared named entities to study short information cascades. We are not interested in long-range, general topics, nor in having to rely on the occurrence of very specific, rare strings. Instead, our goal is to identify topics that can identified by a set of n-grams and a well bounded period of time, and that represent simple, self-contained units of activity.% on the network.

%We propose a rather holistic approach, that takes advantage of three types of data present in the crawl: the textual content of the blog posts, the times at which the blog posts where created and the blogs that the posts originated from.
We propose a rather holistic approach, that takes advantage of both the textual content of  blogs posts, and the times at which these posts where published.

%The process of topic detection we propose consists of a classical sequence of steps:
The process of topic detection we propose consists of a classical sequence of treatments that we perform on our dataset:

\begin{enumerate}
\item Part-of-speech tagging and lemmatisation of each post's title and content in order to enumerate every relevant n-grams in the corpus.
%\item Chunking of text by punctuation marks.
%\item Enumeration of n-grams present in the corpus.
\item Detection and filtering of n-gram temporal bursts.
\item Merging of redundant n-gram bursts into unique topics.
\end{enumerate}

\subsection{linguistic treatment}
We perform the first step using the TreeTagger tool~\cite{Schmid94}. In this step we generate a new version of each posts title and textual content, where each word is lemmatised and augmented with a part-of-speech tag. 

%In the second step, the corpus of text generated by the previous step is divided into chunks, delimited by punctuation marks.

We then divide the corpus of text generated by the previous step into chunks, delimited by punctuation marks.
Afterwords, we find all the n-grams that occur in the chunks produced by the previous step. This search is constrained by a set of rules, as to not generate an intractable amount of n-grams, and explore only cases we believe are likely to lead to meaningful topics.
 The rules are the following:
\begin{itemize}
\item N-grams must have two or more words.
\item An n-gram must contain at lease one noun.
\item All words that are not nouns, verbs, adjectives or numbers are discarded.
\item All n-grams that contain words in a special set called \emph{stop-words list} are rejected.
\end{itemize}These rules are empirical, having been obtained by experimentation with real datasets. The word set in the last rule contains words that have a strong temporal meaning, and that would later on lead to the detection of meaningless temporal bursts of usage. We used a set containing names of months, days of the week and holiday seasons (like Christmas), in both French and English.

\subsection{Temporal bursts detection}

In the second phase, we analyse the pattern of occurrence of each n-gram, dividing the period of observation into bursts of activity. For this purpose, we devised an algorithm that iteratively divides the timeline into intervals, aiming at the maximization of a value we will call the \emph{burst ratio}. Let us consider an ordered set $T = \{t_0, t_1, ..., t_n\}$ (in ascending order), where each element is the time of an occurrence of the n-gram. Furthermore, any two consecutive elements of $T$ must originate from different blogs. This guarantees that a burst can only be maintained by the participation of multiple blogs. 

\begin{figure}
\begin{center}
\includegraphics[scale=0.8]{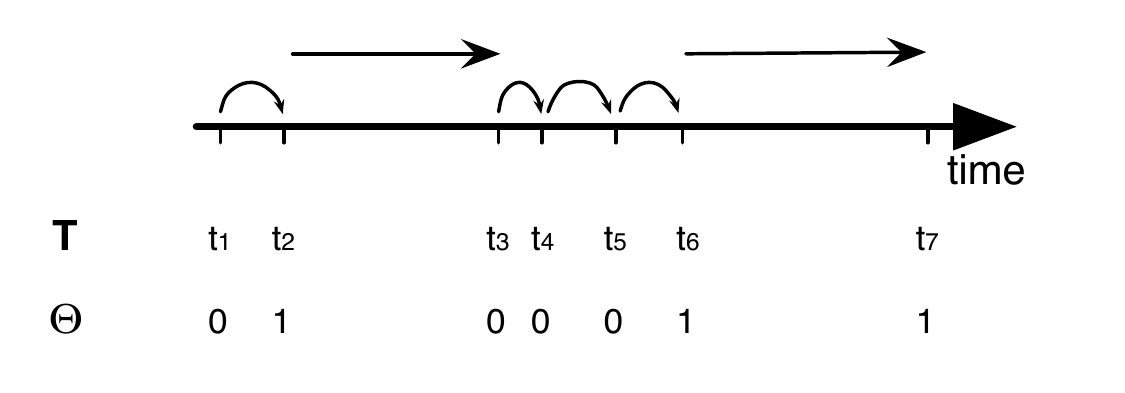} 
\end{center}
\VSP
\caption{Example of a sequence of occurrences of a given ngram. The ordered sets $T$ and $\Theta$ are depicted. Inter-bursts and intra-burst intervals are represented by arrows (respectively straight and curved). }
\VSP
\label{ngrams2topics}
\label{gridbrain}
\end{figure}

We are interested in partitioning $T$ into subsets which correspond to temporal bursts.
Let us consider the ordered set $\Theta = \{\theta_0, \theta_1, ..., \theta_n\}$ where $\theta_k = 1$ if element $t_k$ is the last element of a burst, and $\theta_k = 0$ otherwise. Each time $\theta_{k}$ equals $1$ it means that the burst ends at $t_{k}$. 
Given a partition  $\Theta$ of the sequence of a n-gram into bursts, it is straightforward to compute the time-lag between the end of a burst and the beginning of the next burst or the time-lag between two occurrences inside the same burst. We can compute the average time-lag between two consecutive bursts or the average interval inside each burst on the whole timeline as follows:

%\begin{equation}
%\displaystyle V_{inter} = \begin{cases} 0, & \mbox{if } \sum_{i=1}^{n-1} L_i = 0 \\
%\\
%\displaystyle \frac{\sum_{i=1}^{n-1} (T_{i+1} - T_i) L_i}{\sum_{i=1}^{n-1} L_i}, & \mbox{if } \sum_{i=1}^{n-1} L_i > 0 \end{cases}
%\end{equation}
%
%\begin{equation}
%\displaystyle V_{intra} = \begin{cases} 0, & \mbox{if } \sum_{i=1}^{n-1} L_i = 0 \\
%\\
%\displaystyle \frac{\sum_{i=1}^{n-1} (T_{i+1} - T_i) (1 - L_i)}{\sum_{i=1}^{n-1} (1 - L_i)}, & \mbox{if } \sum_{i=1}^{n-1} L_i > 0 \end{cases}
%\end{equation}

\begin{align}
\nonumber
\displaystyle V_{\longmapsto}(T,\Theta) =& \displaystyle \frac{\sum_{i=1}^{|T|-1} (t_{i+1} - t_i) \theta_i}{\sum_{i=1}^{|T|-1} \theta_i}, \\
&\mbox{if } \sum_{i=1}^{|T|-1} \theta_i > 0, 0 \mbox{ otherwise}
\end{align}

\begin{align}
\nonumber
\displaystyle V_{ \curvearrowright}(T,\Theta) =& \displaystyle \frac{\sum_{i=1}^{|T|-1} (t_{i+1} - t_i) (1-\theta_i)}{\sum_{i=1}^{|T|-1} (1-\theta_i)}, \\
&\mbox{if } \sum_{i=1}^{|T|-1} \theta_i > 0, 0 \mbox{ otherwise}
\end{align}

We also define the minimum inter-burst interval $m_{\longmapsto}(T,\Theta)$ as:
$$m_{\longmapsto}(T,\Theta)={min}_{\{i<|T|,\theta_{i}=1\}}(t_{i+1}-t_{i})$$

We then define the \emph{burst ratio}, $\rho(T,\Theta)$ as:
%
%\begin{equation}
%\displaystyle R_b = \begin{cases} 0, & \mbox{if } V_{intra} = 0 \\
%\\
%\displaystyle \frac{V_{inter}}{V_{intra}}, & \mbox{if } V_{intra} > 0 \end{cases}
%\end{equation}
%
%$$\displaystyle R_b = 
%\displaystyle \frac{V_{inter}}{V_{intra}},  \mbox{if } V_{intra} > 0\mbox{ , }0\mbox{ otherwise } 
%$$
%

$$\displaystyle \rho(T,\Theta) = 
\displaystyle \frac{V_{\longmapsto}(T,\Theta)}{V_{\curvearrowright}(T,\Theta)},  \mbox{if } V_{\curvearrowright}(T,\Theta) > 0\mbox{ , }0\mbox{ otherwise } 
$$

Simply put, $\rho(T,\Theta)$ is the ratio of the mean time interval between bursts to the mean time interval between elements inside bursts.

On algorithm~\ref{clustering_algorithm} we present the pseudo-code that describes the clustering method. The process is started with all the elements of $\Theta$ initialized to $0$, meaning that in the initial state, all n-gram occurrences are considered to belong to a single burst. The algorithm iteratively tries to add new divisions to $\Theta$, keeping the ones that increase the \emph{burst ratio}, until no further improvement is possible.

\begin{algorithm}
\caption{Pseudo-code of algorithm to perform temporal clustering of n-gram occurrences into bursts.}
\label{clustering_algorithm}
\begin{algorithmic}
{\footnotesize \STATE $stop \gets False$
\WHILE {$stop = False$}
    \STATE $best\_burst\_ratio \gets -1$
    \STATE $best\_postion \gets -1$
    \FOR{$pos=1$ to $|T|$}
        \IF {$\Theta\_{pos} = 0$}
            \STATE $aux\_\Theta \gets \Theta$
            \STATE $\Theta_{pos} \gets 1$
            \STATE $burst\_ratio \gets \rho(T, aux\_\Theta)$
            \STATE $min\_inter\_interval \gets m_{\longmapsto}(T, aux\_\Theta)$
            \IF {$burst\_ratio < \alpha$ \textbf{or} $min\_inter\_interval < \beta$}
                \STATE $burst\_ratio \gets 0$
            \ENDIF
            \IF {$burst\_ratio > best\_burst\_ratio$}
                \STATE $best\_burst\_ratio \gets burst\_ratio$
                \STATE $best\_pos \gets pos$
            \ENDIF
        \ENDIF
    \ENDFOR
    \IF {$best\_pos > 0$}
        \STATE $\Theta_{best\_pos} \gets 1$
    \ELSE
        \STATE $stop \gets True$
    \ENDIF
\ENDWHILE}
\end{algorithmic}
\end{algorithm}

Parameters $\alpha$ and $\beta$ determine, respectively, the minimum \emph{burst ratio} and interval between bursts (in days) that are accepted. These parameters allow us to prevent the formation of bursts that are not sufficiently separated, both in relation to the average interval between n-gram occurrences and in absolute value. For our purposes, we experimentally determined $\alpha = 5$ and $\beta = 5$ to be good values.

We devised our own burst detection algorithm instead of using one of the available ones, due to the specific requirements of our approach. For example, the weighted automaton model described in~\cite{klei:burs} is very suitable for detecting bursts at quantifiable levels of intensity, but does not lend itself to the detection of bursts with well defined limits. For the probabilistic model we are going to describe in the following section, it is crucial that we consider bursts with well defined limits, as not to lose initial or late arrivals. Our algorithm detects cases where the activity on a certain n-gram set can be characterized by intervals with a sufficient level of activity, separated by large enough intervals of no activity. 

Finally we filter the n-gram bursts, only accepting the ones that meet the following criteria:

\begin{itemize}
\item A minimum number of blogs participating in the burst of 4.
\item A minimum average time between posts participating in the burst of 1 hour.
\item A maximum average time between posts participating in the burst of 1 day.
\item A minimum burst duration of 3 days.
\item A maximum total duration of all the bursts of the n-gram of 1 month.
\end{itemize}

The purpose of these rules is to end up with n-gram bursts that are more likely related to a real topic. We discard bursts that are too sparse, too dense, too short lived or defined by an n-gram that is too common.

\subsection{Merging n-gram bursts into topics} 

Finally, on the last phase, we extract a set of topics from the set of n-gram bursts that resulted from the previous step. We define a topic as a tuple $(\{g_0, g_1, ..., g_n\}, t, t')$, consisting of a set of n-grams occurring between times $t$ and $t'$. Topics are defined with the minimum possible set of n-grams for maximum generality. Figure~\ref{ngrams2topics} illustrates on a real example how the n-gram bursts are selected to define a topic. The underlying idea is the following: consider two n-gram bursts, defined by n-grams $g_a$ and $g_b$, occurring over time intervals $[t_a, t'_a]$ and $[t_b, t'_b]$. Furthermore, consider that the sequence of words in n-gram $g_b$ is a sub-sequence of the sequence of words in n-gram $g_a$, and that $t_a \ge t_b$ and $t'_a \le t'_b$. Referring to figure~\ref{ngrams2topics}, this could be exemplified by $g_a= ``r\acute{e}gion\ avoir\ apporter\ contribution\ d\acute{e}bat''$ and $g_b= ``apporter\ contribution\ d\acute{e}bat''$. We assume that, in this kind of situation, it is very likely that both bursts belong to the same topic. $g_b$ is more general than $g_a$, because it includes all the cases covered by $g_a$, while the opposite is not necessarily true.

We transverse the entire set of n-gram bursts, in descending order of the number of words contained in their n-gram. For each burst, we look for bursts ahead in the set with n-grams that are a sub-sequence of the first one, and with time intervals that contain the interval of the first one. If such bursts are found, the original burst is discarded. If one of the bursts found is already assigned to a topic, we also assign the other bursts found to that topic, otherwise we assign all bursts found to a new topic.

\begin{figure}
\begin{center}
\includegraphics[scale=0.5]{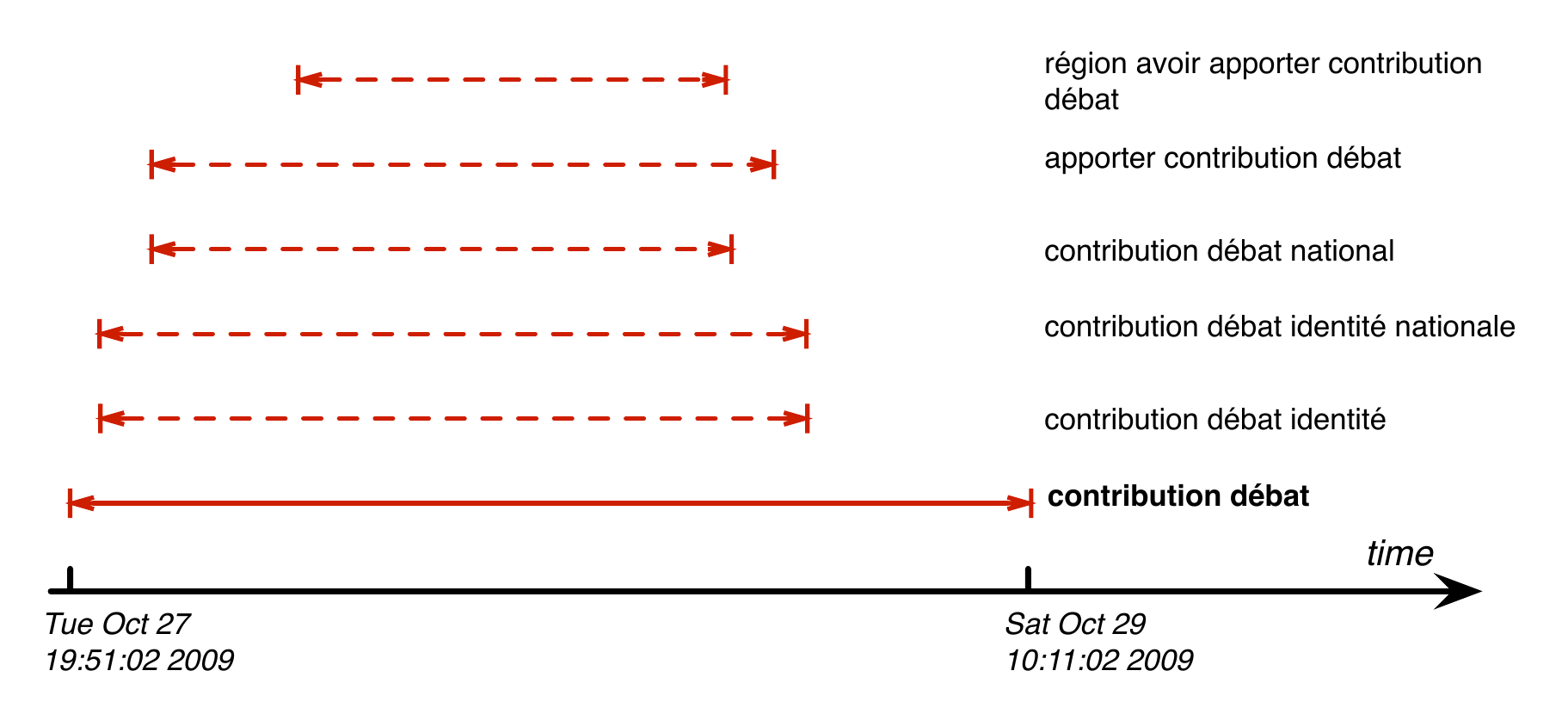} 
\end{center}
\VSP
\caption{Example of selection of n-gram bursts to define a topic. Bursts in sold line are selected for the topic definition, while bursts in dashed lines are discarded.}
\VSP
\label{ngrams2topics}
\label{gridbrain}
\end{figure}

%% file: soccom10-probabilistic.tex
\section{Probabilistic Precedence Scoring}
\label{sec4}
After the process described in the previous section, we now have a set of topics, and know which blogs participated in each topic and at what time. We are now in the position of defining a probabilistic model that estimates the tendency that blogs have to participate in topics before other blogs.

We will start by defining a \emph{dyadic precursor score} from blog $b$ to blog $b'$. We will call this score $\gamma(b, b')$. Let us define $A$ as the set of all topics where both blogs participate, and $Y$ as the subset of $A$ where the first participation of $b$ precedes the first participation of $b'$. We also define $C$ as a vector of probabilities. Each element of $C$ is the probability that $b$ participates on a topic before $b'$ by chance. We will detail later how these probabilities are computed. We know define the likelihood of $\gamma(b, b')=p$, given $A$, $Y$ and $C$:

\begin{align}
\label{likelihood1}
\nonumber
& \lambda(\gamma(b, b')=p|A,Y,C) = \\
& \sum_{\substack{Z \cup R=Y \\
Z \cap R = \emptyset}} \lambda(\gamma(b, b')=p|A,Y,C,Z,R)
\end{align}

The likelihood in equation~\ref{likelihood1} is defined as the sum of the likelihoods for all possible hypothesis of the appearances of $b$ before $b'$ being caused by a temporal relationship or by chance. 
The set $Y$ of topics where the first participation of $b$ precedes the first participation of $b'$ can be decomposed as the union of the set $Z$ of topics where $b$ is assumed to display a behavior of precedence over $b'$, and the set $R$ of topics where $b$ is assumed to precede $b'$ by chance.
%For each hypothesis, $Z$ is the set of topics where $b$ is assumed to display a behavior of precedence over $b'$, and $R$ is the set of topics where $b$ is assumed to precede $b'$ by chance. 
We define the likelihood of each hypothesis as:

\begin{equation}
\lambda(\gamma(b, b')=p|A,Y,C,Z,R) = P_Z(A, Z, p) \cdot P_R(A, R, C)
\end{equation}

$P_Z(A,Z,p)$ is the probability that $b$ precedes $b'$ in the topics in $Z$ and not in the topics in $A \setminus Z$, given a probability of a precedence relationship of $b$ over $b'$ of $p$. $P_R(A, R, C)$ is the probability that $b$ precedes $b'$ by chance for the topics in $R$, and not for the topics in $A \setminus R$, given $C$. These probabilities are defined as:
%$P_Z(T, L, p)$ is the probability that $b$ precedes $b'$ in the topics in $L$ and not in the topics in $T \setminus I$, given a probability of a precedence relationship of $b$ over $b'$ of $p$. $P_R(T, R, C)$ is the probability that $b$ precedes $b'$ by chance for the topics in $R$, and not for the topics in $T \setminus R$, given $C$. These probabilities are defined as:

\begin{equation}
P_Z(A, Z, p) = p^{|Z|} (1 - p)^{|A| - |Z|}
\end{equation}

\begin{equation}
P_R(A, R, C) = \prod_{r \in R} C_r \prod_{r \in A \setminus R} 1 - C_r
\end{equation}

Now we have to define how to compute the probabilities $C_{r}$ that topic $r$ is mentioned by $b$ before $b'$. We compute these probabilities by taking into account the total number of posts published by each blog during the time interval of the topic, in the following way:

\begin{equation}
C_r = \frac{Np(b, [t_s(r); t_e(r)])}{Np(b, [t_s(r); t_e(r)]) + Np(b', [t_s(r); t_e(r)])}
\end{equation}

%\begin{equation}
%C_r = \frac{Np(b, T_S(r), T_E(r))}{Np(b, T_S(r), T_E(r)) + Np(b', T_S(r), T_E(r))}
%\end{equation}

$t_s(r)$ is the time of the beginning of topic $r$ and $t_e(r)$ is the time of its end. $Np(j, t, t')$ gives the number of posts published by blog $j$ between times $t$ and $t'$. Simply, this expression reflects the idea that, the higher the number of posts of blog $b$ as compared to the total number of posts from both blogs in the time interval, the more likely $b$ is to publish the first post on the topic by chance. We do not consider the overall posting rates of the blogs, as these change over time.

%\TBR{Also, considering the number of posts (related to the topic or not) published during the time interval of the topic lets us take into account the effect of the topic itself on the posting rates of the blogs.}

The computation of the likelihood expressed in~\ref{likelihood1} suffers from combinatorial explosion. In fact, the number of computations that have to be performed to calculate $\lambda(\gamma(b, b')=p|A,Y,C,Z,R)$ scales exponentially with $|Y|$. For this reason, when $|Y|$ is above $15$, we resort to an estimation based on sampling.

%\begin{equation}
%N_{comps} = \sum_{i=0}^{|Y|} \binom{|Y|}{i}
%\end{equation}
%}

Finally, we estimate $\gamma(b, b')$ by calculating the mean of the possible values it can take ($\gamma(b, b') \to [0, 1]$), weighted by their likelihood:

\begin{equation}
\label{L_mean}
\gamma(b, b') = \frac{\int_0^1 l(\gamma(b, b')=p|A,Y,C) \cdot p \cdot dp}{\int_0^1 l(\gamma(b, b')=p|A,Y,C) \cdot dp}
\end{equation}

Not having an analytical solution for equation~\ref{L_mean}, we use Monte Carlo integration.

Having a way to compute dyadic precursor scores, we are now interested in scoring the blogs according to their overall precursor/laggard behaviors over the entire network. For this purpose, we will define two metrics: the global precursor score ($P$) and the laggard score ($L$).

A dyadic precursor score $\gamma(b, b')$ can be interpreted as the probability that a post from blog $b'$ participates in a topic under a temporal relationship with blog $b$, where $b$ precedes $b'$, given that both blogs are known to participate in that topic. We can remove the topic co-participation assumption using Bayes' theorem. Considering $M$ to be the event of the post participating in the topic under the temporal relationship, and $H$ to be the event of the post for blog $b'$ participating in a topic where blog $b$ also participates:

\begin{equation}
\gamma(b, b') = P_r(M|H)
\end{equation}

\begin{equation}
P_r(M|H) = \frac{P_r(H|M) P_r(M)}{P_r(H)}
\end{equation}

\begin{equation}
\omega(b,b') = P_r(M) = P_r(M|H) P_r(H) = \gamma(b,b') P_r(H)
\end{equation}

We will call $\omega(b,b')$ the adjusted dyadic precursor score. Notice that $P_r(H|M) = 1$, because if the post participates in a topic under a temporal relationship with the other blog, the blogs will necessary co-participate in that topic.

We define the global precursor score for a blog $b$ ($P(b)$) as the mean of all adjusted dyadic precursor scores where $b$ is the origin, and the laggard score ($L(b)$) as the mean of all adjusted dyadic precursor scores where $b$ is the target. Being $B$ the set of all blogs in the network:

\begin{equation}
P(b) = \frac{1}{|B|-1}\sum_{b' \in B \setminus \{b\}} \omega(b, b')
\end{equation}

\begin{equation}
L(b) = \frac{1}{|B|-1}\sum_{b' \in B \setminus \{b\}} \omega(b', b)
\end{equation}

%% file: soccom10-discussion.tex
\section{Results and Discussion}

\label{sec5}

The protocol described in the previous sections was applied to a dataset generated from a crawl of the French  political blogosphere, consisting of $916$ blogs, between the days of October 1$^{st}$ 2009 and February 11$^{th}$ 2010. During this period, $40,191$ posts were published, containing $16,909$ citation links to other blogs in the network. 
We applied our topic detection process on this data and identified   $2,619$ different topics. 

We then computed the global precursor and laggard scores according to the process described in the previous section for each blog that published at least $7$ posts during the whole observation period.  We discarded nearly $300$
%291
 blogs with very low posting rates because of the noise they may introduce into the computation of the global scores.

%The protocol described in the previous sections was applied to a data set generated from a crawl of the French  political blogs, consisting of 916 blogs, between the days of October 1$^{st}$ 2009 and February 11$^{th}$ 2010. During this period, 40191 posts were published, containing 16909 links to other blogs in the network.
%
%The application of the topic detection process previously described resulted in the identification of 2619 topics. 661 of the 916 blogs (72\%) participated in an least one of these topics.
%
%Dyadic leader scores were computed for all pairs of blogs in the network. From this point on, we decided to only consider blogs with at least seven posts during the observation period. We found that blogs with very low post counts tend to introduce considerable noise into the global scores. This is due to their small samples, which can lead to distorted $\gamma$ scores. The number of blogs in the network with at least seven posts during the observation period is 625. The global precursor and laggard scores were computed for each blog in this reduced set, according to the process described in the previous section. All following results are based on the reduced set.

\begin{figure}
\begin{center}
\includegraphics[scale=0.45]{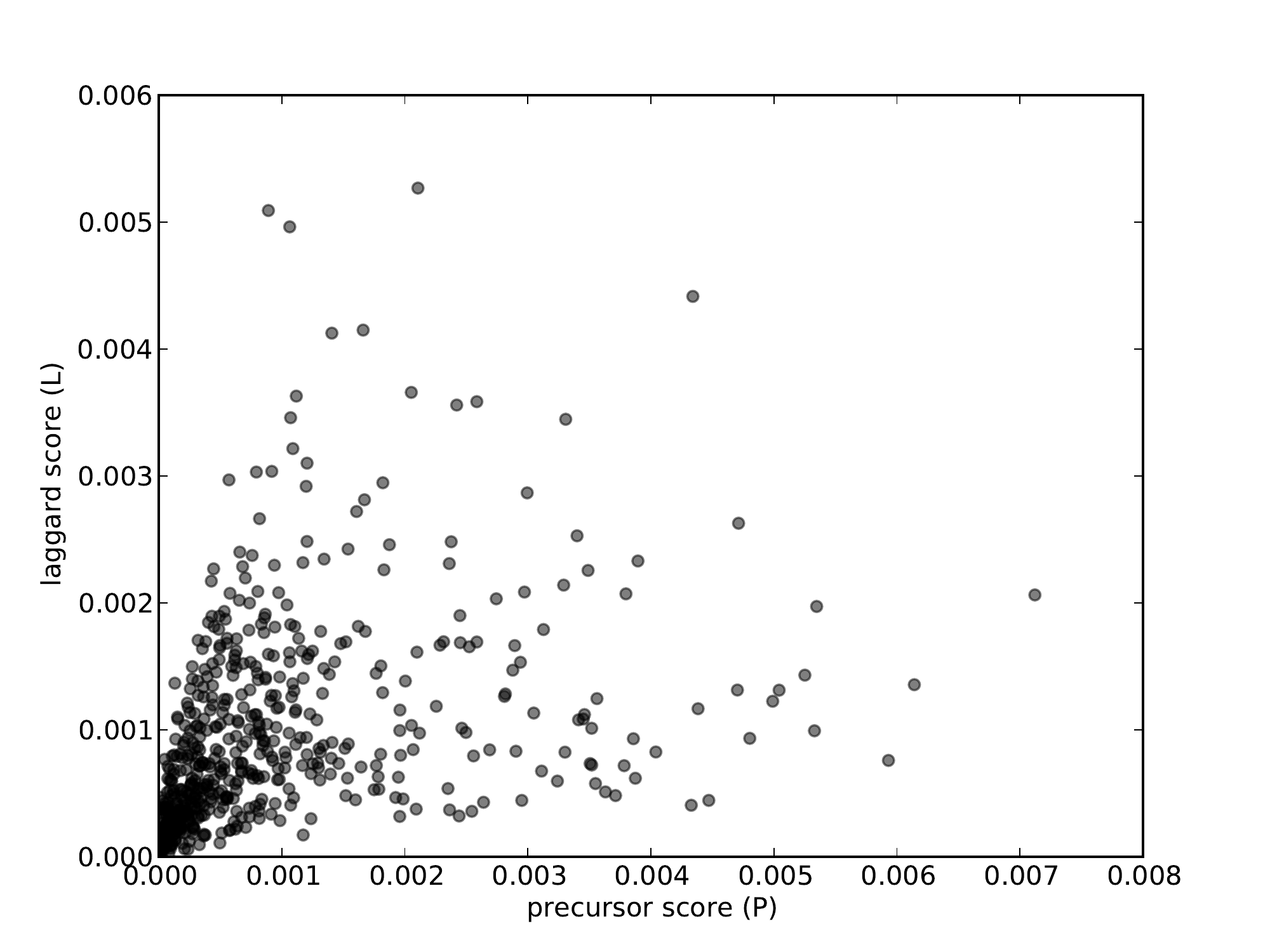} 
\end{center}
\VSP
\caption{Scatter plot of precursor ($P$) vs. laggard ($L$) scores for all blogs in the network.}
\VSP
\label{pre_lag}
\end{figure}

Figure~\ref{pre_lag} shows a scatter plot of the blogs, positioned in the plane according to their global precursor and laggard scores: $P$ and $L$. This plot gives us an overview of the structure of the network in terms of precursor/laggard behaviors. It can be observed that there is a dense cluster of blogs near the origin, with the distribution of blogs rarefying in both the $x$ and $y$ directions.

A blog may be situated in the low scores cluster for different reasons. It could be that it does not tend to participate in popular topics (which also means that the topics it discusses are not spread through the network), or it could be that it maintains relationships of influence with other blogs which are close to being symmetrical. This type of relationship between two blogs makes it approximately equally likely that each blog influences the other to enter a topic. Our scores are not capable of distinguishing a symmetrical influence relationship from an indirect relationship\footnote{Since the blog network is not a closed system, two blogs could have a very similar set of external influences, leading to the same temporal patterns they would display if influencing each other in a symmetrical way.}.
%A blog may be situated in the low scores cluster for different reasons. It could be that it does not tend to participate in popular topics (which also means that the topics it discusses are not spread through the network), or it could be that it maintains relationships of influence with other blogs which are close to being symmetrical. This type of relationship between two blogs makes it approximately equally likely that each blog influences the other to enter a topic. Our scores are not capable of distinguishing a symmetrical influence relationship from an inexistent relationship, nor do we believe that it is possible to establish that distinction just from analyzing the crawl data. Since the blog network is not a closed system, two blogs could have a very similar set of external influences, leading to the same temporal patterns they would display if influencing each other in a symmetrical way.

In the study of blog networks, it is common to establish popularity metrics based on the URL links that point to a blog.  We compute the in-degree of a blog as the number of blogs that link to it at least once during the observation period, as well as the classical page rank. Our goal is to compare those metrics based on the topology of the hyperlinks network with our temporal semantic based scores.  

%\TB{I erased a large portion of text here}
%
%In the study of blog networks, it is common to establish popularity metrics based on the URL links that point to a blog. We compared this type of metric with our temporal semantic scores. We define the link in-degree of a blog as the number of blogs that link to it, at least one time, during the observation period. We opted to disregard factors links the number of posts from a blog that point to another, or even the total number of links, to avoid distortions caused by more complex behaviors, like blogs with an unusual high tendency for linking to other specific blogs or to other blogs in general. Our goal is to unveil network-level phenomena.

\begin{figure}[!bh]
\begin{center}
\includegraphics[scale=0.45]{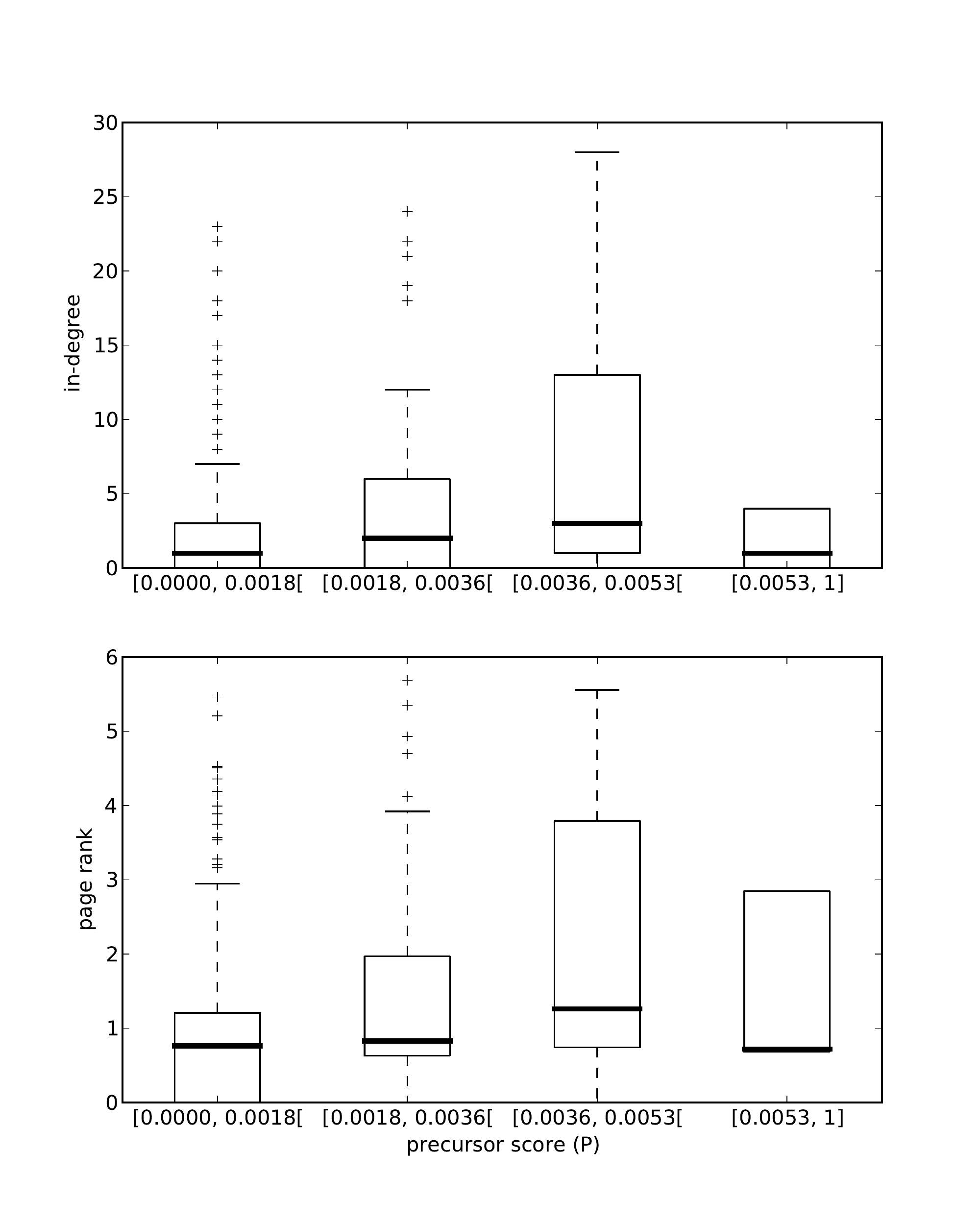} 
\end{center}
\VSP
\caption{Above: box plots of in-linking distributions for intervals of precursor scores. Below: box plots of page rank distributions for intervals of precursor scores.}
\VSP
\label{pre_indegree_pr}
\end{figure}

Figure~\ref{pre_indegree_pr} shows box plots of in-linking and page rank per interval of precursor score. The two plots present similar shapes, showing an increase in both in-link degrees and page ranks up to the third bar. On the fourth bar there is a clear decrease, suggesting that the precursor behavior is positively correlated with blog popularity only up to a certain point.
%Figure~\ref{pre_indegree_pr} shows box plots of in-linking and page rank per interval of precursor score. The two plots present similar shapes, showing an increase in both in-link degrees and page ranks up to the third bar. On the forth bar there is a clear decrease, suggesting that the precursor behavior has a positive effect in link popularity only up to a certain point.

%\includegraphics[scale=0.45]{figs/pc_ind.pdf} 
%\end{center}
%\caption{Log-lin plot of precursor score vs. mean post count on the left, log-lin plot of post count vs. mean in-degree on the right.}
%\label{pc_ind}
%\end{figure}

%On the plots shown on figure~\ref{pc_ind}, we can see that the mean post count increases with the precursor score, and also that the mean in-degree increases with the post count. This contrasts with the previous finding that the mean in-degree does not always increase with the precursor score. Blogs with higher precursor scores tend to post more, but this does not translate to the increased in-link degree we observe when considering the entire network.

\begin{figure}
\begin{center}
\includegraphics[scale=0.45]{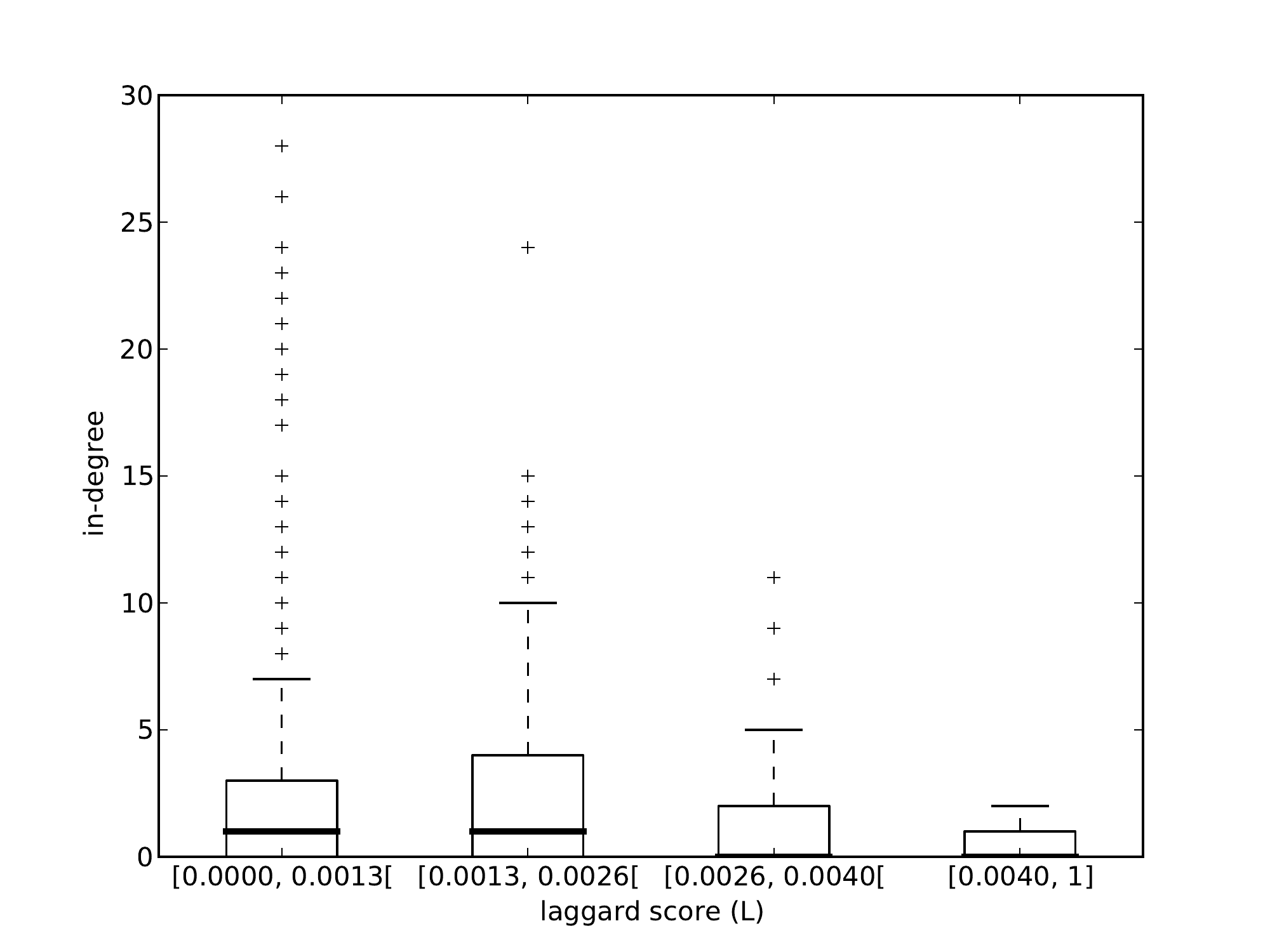} 
\end{center}
\VSP
\caption{Box plots of in-linking distributions for intervals of laggard scores.}
\VSP
\label{lag_indegree}
\end{figure}

In figure~\ref{lag_indegree} we plot in-linking per interval of laggard score. This plot is more noisy and the pattern is less clear than the previous one% in figure~\ref{pre_indegree_pr}
. Higher laggard scores appear to have a detrimental effect on link popularity. Although not shown, a similar pattern was found when comparing page ranks to laggard scores.
%In figure~\ref{lag_indegree} we plot in-linking per interval of laggard score. This plot is more noisy and the pattern is less clear than the own shown in figure~\ref{pre_indegree_pr}. Higher laggard scores appear to have a detrimental effect on link popularity. Although not shown, a similar pattern was found when comparing page ranks to laggard scores.

%\begin{table}[!t]
%% increase table row spacing, adjust to taste
%\renewcommand{\arraystretch}{1.3}
% if using array.sty, it might be a good idea to tweak the value of
% \extrarowheight as needed to properly center the text within the cells
%\caption{Mean in-degree relationships for classes of blogs determined according to precursor and laggard score intervals.}
%\label{table_high_low}
%\centering
%\begin{tabular}{|c|c|c|c|c|}
%\hline
%   & \emph{lp/ll} & \emph{lp/HL} & \emph{HP/ll} \\
%   & (2.08)       & (1.59)       & (6.19) \\
%\hline
%\emph{lp/HL} & \textbf{lp/HL $<$ lp/ll *} & - & - \\
%(1.59) &  &  &  \\
%\hline
%\emph{HP/ll} & \textbf{HP/ll $>$ lp/ll **} & \textbf{HP/ll $>$ lp/HL ***} & - \\
%(6.19) &  &  &  \\
%\hline
%\emph{HP/HL} & HP/HL $>$ lp/ll & \textbf{HP/HL $>$ lp/HL ***} & HP/HL $<$ HP/ll \\
%(3.50) &  &  &  \\
%\hline
%\end{tabular}
%\end{table}

\begin{table}[!t]
%% increase table row spacing, adjust to taste
\renewcommand{\arraystretch}{2.0}
% if using array.sty, it might be a good idea to tweak the value of
% \extrarowheight as needed to properly center the text within the cells
\caption{Significance of mean in-degree relationships for classes of blogs determined according to precursor and laggard score intervals.}
\VSP
\label{table_high_low}
\begin{center}
\begin{tabular}{cc|c|c|c|c|}
\cline{3-6}
& & $2.08$ & $6.19$ & $1.59$ & $3.50$ \\ %\cline{3-6}
& & $pl$ & $\textbf{P}l$ & $p\textbf{L}$ & $\textbf{PL}$ \\ \hline%\cline{1-6}
%\multicolumn{1}{|c|}{\multirow{2}{*}{Powers}} &
\multicolumn{1}{|c}{$2.08$} & \multicolumn{1}{c|}{$pl$} &\multicolumn{1}{c}{ \cellcolor[gray]{0.9}} & \multicolumn{1}{c}{ \cellcolor[gray]{0.9}} & \multicolumn{1}{c}{ \cellcolor[gray]{0.9}} &\multicolumn{1}{c|}{  \cellcolor[gray]{0.9}}      \\ \cline{1-2}

\multicolumn{1}{|c}{$6.19$} & \multicolumn{1}{c|}{$\textbf{P}l$} & \multicolumn{1}{c}{  \textbf{**} }& \multicolumn{1}{c}{ \cellcolor[gray]{0.9}} & \multicolumn{1}{c}{ \cellcolor[gray]{0.9}} &  \cellcolor[gray]{0.9}     \\ \cline{1-2}

\multicolumn{1}{|c}{$1.59$} & \multicolumn{1}{c|}{$p\textbf{L}$} & \multicolumn{1}{c}{ \textbf{*}} & \multicolumn{1}{c}{ \textbf{***}} & \multicolumn{1}{c}{ \cellcolor[gray]{0.9}} & \multicolumn{1}{c|}{ \cellcolor[gray]{0.9}}      \\ \cline{1-2}

\multicolumn{1}{|c}{$3.50$} & \multicolumn{1}{c|}{$\textbf{PL}$} &\multicolumn{1}{c}{ }  & \multicolumn{1}{c}{}   & \multicolumn{1}{c}{ \textbf{***}} & \multicolumn{1}{c|}{ \cellcolor[gray]{0.9} }     \\ \cline{1-6}
\end{tabular}
\VSP
\end{center}
\end{table}

In order to derive general principles, we divided the blog set into four classes. Each class is characterized by a high or low precursor score and a high or low laggard score. A precursor score is considered low if it is equal or lesser than the mean precursor score for the entire set ($P \in [0, \overline{P}[$), and high otherwise ($P \in ]\overline{P}, 1]$). Laggard scores are classified in an analogous fashion. We use the notation $p$ for low precursor, $\textbf{P}$ for high precursor and so on. The class $\textbf{P}l$, for example, is the one containing blogs with an high precursor score and low laggard score.

In each cell of table~\ref{table_high_low} we perform a comparison between the mean in-link degree of each class. The statistical significance of the differences was determined using \emph{Wilcoxon} rank sum tests. We use a number of $*$ symbols to denote the level of significance found. One $*$ if $p-value < 0.05$, two if $p-value < 0.01$ and three if $p-value < 0.001$. The mean in-degrees for classes are shown in row and column headers.

When comparing the two classes with low laggard scores, the one with an high precursor score has a higher mean in-degree. The same is true of the two classes with high laggard scores. When comparing the two classes with a low precursor score, the one with the low laggard score has the higher mean in-degree. In the two cases where no significance was found, the p-value was very close to $0.05$, suggesting that the relationships are likely true, but we have insufficient data to be certain. This confirms that higher precursor scores and lower laggard scores have a positive effect on in-linking. These results also show that the two scores are not just reflecting the effect of participating in discussions. In fact, both scores require higher participation for higher values, but have opposite effects.
%Generally, we observe that higher precursor scores and lower laggard scores have a positive effect on in-linking. These results also show that the two scores are not just reflecting the effect of participating in discussions. In fact, both scores require higher participation for higher values, but have opposite effects.

It is clear, however, that these general principles do not tell the whole story. The box plots show that, despite the general principles, blogs with high precursor scores are not necessarily rewarded with high in-link degrees. 

\begin{figure}
\begin{center}
\includegraphics[scale=0.45]{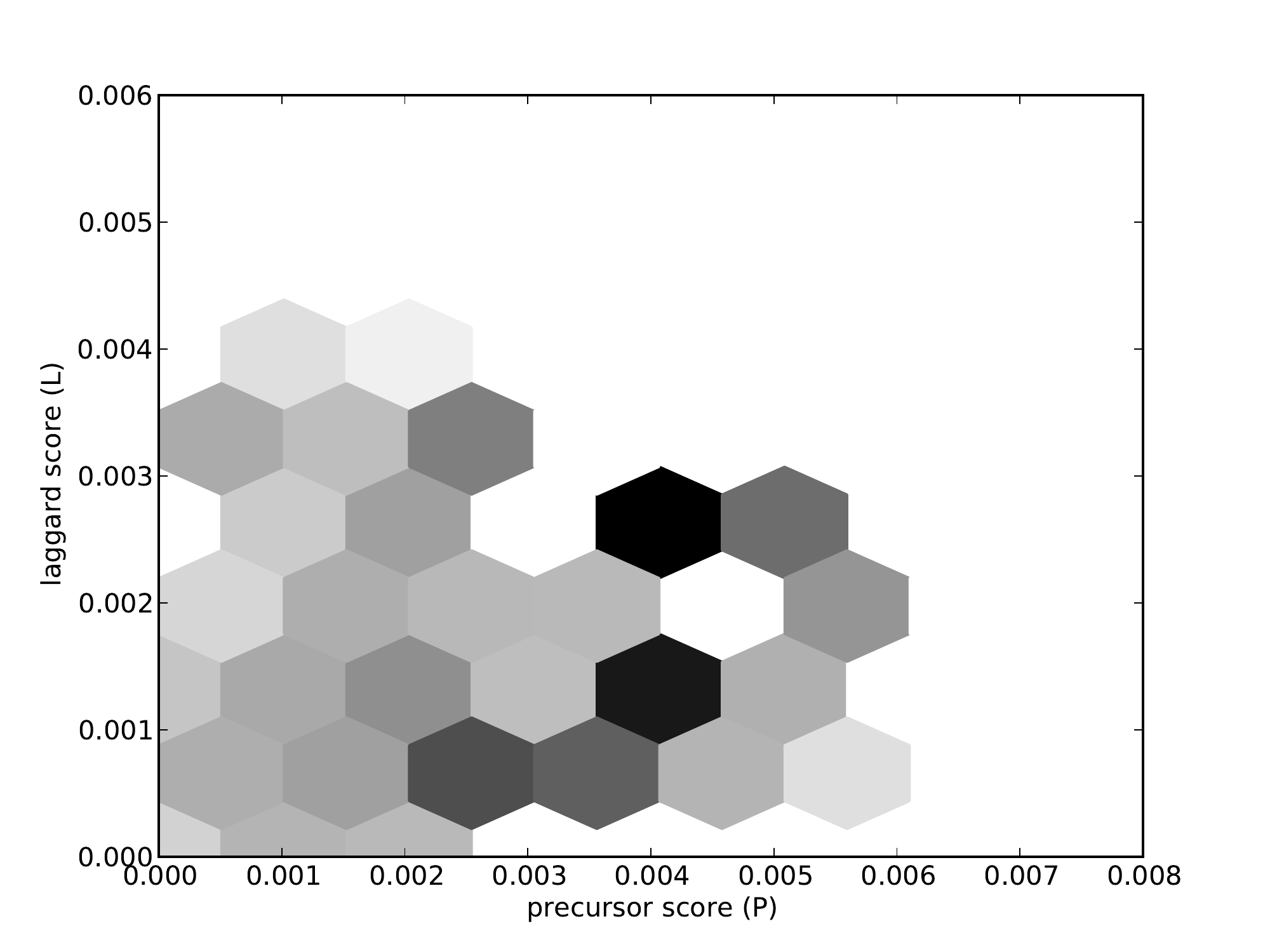} 
\end{center}
\VSP
\caption{Hexagonal binning plot displaying mean in-linking per region of precursor and laggard scores. The darker the color, the higher the in-linking.}
\VSP
\label{hexbin}
\end{figure}

This becomes more obvious by observing the hexagonal binning plot, shown in figure~\ref{hexbin}.  It displays the mean in-linking per region of precursor and laggard scores. The darker the color, the higher the in-linking mean. It clearly confirms for example that higher precursor score does not guarantee higher in-degree. 
%This fact becomes more obvious by observing the scatter plot in figure~\ref{scatter_pre_indegree}. For a more detailed view, we generated an hexagonal binning plot, shown in figure~\ref{hexbin}. It displays the mean in-linking per region of precursor and laggard scores. The darker the color, the higher the in-linking mean.

%\begin{figure}
%\begin{center}
%\includegraphics[scale=0.45]{figs/scatter_pre_indegree.pdf} 
%\end{center}
%\caption{Scatter plot of precursor score vs. in-degree for all blogs in the network.}
%\label{scatter_pre_indegree}
%\end{figure}

To validate our protocol and experimental results, we generated four lists of ten blogs. We determined the position of each blog on a plane, where dimension $x$ is the precursor score, and $y$ the link in-degree. Both axis were converted to a logarithmic scale and normalized to $[0, 1]$ intervals.
From this spatial distribution, list 1 contains the blogs closest to point $(0, 0)$ - low precursors, low in-degree; list 2 the blogs closest to $(0, 1)$ - low precursors, high in-degree; list 3 the blogs closest to $(1, 0)$ - high precursors, low in-degree and list 4 the blogs closest to $(1, 1)$ - high precursors, high in-degree.

We then provided these four lists to an expert on the French blogosphere. She had no prior knowledge of our classification process. We simply asked her if she could notice  any significant pattern inside  groups.
She described blogs of list 1, which belong to the category of low precursor and low in-degree, as very ``small'' blogs essentially concerned with regional or local issues. According to her, list 2 (low precursors, high in-degree) is typically composed of experienced bloggers who emerged during the last presidential election in 2008 and now gather together despite their political differences. As such their pattern of linking is similar to  a ``rich-club'' which  may explain their high in-degree in spite of their low precursor score. Blogs which have high precursor score and low in-degree (list 3) are exclusively made of copycats. These sites are basically systematically relaying the media  or making reviews of regular papers on  the web. The presence of such behavior in the dataset incidentally explains the sharp decline of mean  in-degree and page rank among blogs with highest precursor scores that we observed previously (Fig. \ref{pre_indegree_pr}). %These kinds of sites  explain the  decrease of mean in-degree and page rank among blogs with the highest precursor score.   
The fourth list is composed of high precursors and high in-degree blogs. All of them have been described  by the expert as very active in political contestation, both from the left and the extreme right,  against the government policy and, more broadly, against the current political balance.

% This  that may lead to very interesting applications

% It appears that some blogs have original content but are more dedicated to broadcasting specific information than to animate chats in the blogosphere. This category of blogs may indeed exert influence other the themes discussed in other blogs but their nature does not encourage these bloggers to cite them. 
%Others only relay information from media, they may have high leadership scores only because they react fast to public events but not because they are read and followed. 

%% file: soccom10-conclusion.tex
\section{Conclusions}

In this work, we strived to extract quantifiable metrics from the wealth of semantic information contained in blogs. We presented a method for the detection of bursts of activity at the semantic level, that was tested on a real data set and shown capable of identifying topics characterized by n-grams and time intervals. We then described a probabilistic model to quantify temporal relationships between blogs. Dyadic precursor scores are able to quantify temporal relationships between pairs of blogs, where one tends to enter a topic before the other, discounting the effects of asymmetrical posting rates. From these dyadic scores we derived two scores to classify blogs according to their overall precursor and laggard behaviors.
%In the work presented in this paper, we strived to to extract quantifiable metrics from the wealth of semantic information contained in blogs. We presented a method for the detection of bursts of activity at the semantic level, that was tested on a real data set and shown capable of identifying topics characterized by n-grams and time intervals. We then described a probabilistic model to quantify temporal relationships between blogs. Dyadic precursor scores are able to quantify temporal relationships between pairs of blogs, where one tends to enter a topic before the other, discounting the effects of asymmetrical posting rates. From these dyadic scores we derived two scores to classify blogs according to their overall precursor and laggard behaviors.

The comparison of these semantic temporal metrics with the more traditional in-link degree based popularity metrics revealed non-trivial relationships between the two. The expert assessment indicates that the scores we proposed lead to relevant distinctions that could not be derived from classical structural based methods only. Search engine ranking algorithms, like the well-known PageRank~\cite{Page1998} used by Google, are more sophisticated than simple reliance on URL link in-degrees. However, they are still based on structural aspects of the web, deriving their estimations from the analysis of the network of URL links.  We found that the precursor/laggard scores are able to identify blogs that have a high tendency to be precursors in topics under discussion, but that would likely not be distinguishable from other blogs with similar page ranks or in-degrees by relying only on this later type of metric. It is conceivable that search engine ranking algorithms could be improved with the approach we propose. Including precursor scores in ranking metrics could help improve the quality of searches, for example the ones related to time sensitive events. It could also reward blogs that generate influential content, but that are not especially popular in the sense of receiving many in-links.

%% file: socialcom2010.bbl
\begin{thebibliography}{10}
\providecommand{\url}[1]{#1}
\csname url@samestyle\endcsname
\providecommand{\newblock}{\relax}
\providecommand{\bibinfo}[2]{#2}
\providecommand{\BIBentrySTDinterwordspacing}{\spaceskip=0pt\relax}
\providecommand{\BIBentryALTinterwordstretchfactor}{4}
\providecommand{\BIBentryALTinterwordspacing}{\spaceskip=\fontdimen2\font plus
\BIBentryALTinterwordstretchfactor\fontdimen3\font minus
  \fontdimen4\font\relax}
\providecommand{\BIBforeignlanguage}[2]{{%
\expandafter\ifx\csname l@#1\endcsname\relax
\typeout{** WARNING: IEEEtran.bst: No hyphenation pattern has been}%
\typeout{** loaded for the language `#1'. Using the pattern for}%
\typeout{** the default language instead.}%
\else
\language=\csname l@#1\endcsname
\fi
#2}}
\providecommand{\BIBdecl}{\relax}
\BIBdecl

\bibitem{lenc:cult}
G.~Lenclud, ``La culture s'attrape-t-elle?'' \emph{Communications}, vol.~66,
  pp. 165---183, 1998.

\bibitem{klei:burs}
J.~Kleinberg, ``Bursty and hierarchical structure in streams,'' in \emph{Proc.
  8th ACM SIGKDD Intl. Conf. on Knowledge discovery and data mining}.\hskip 1em
  plus 0.5em minus 0.4em\relax New York, NY, USA: ACM, 2002, pp. 91--101.

\bibitem{gins:dete}
J.~Ginsberg, M.~H. Mohebbi, R.~S. Patel, L.~Brammer, M.~S. Smolinski, and
  L.~Brilliant, ``Detecting influenza epidemics using search engine query
  data,'' \emph{Nature}, vol. 457, pp. 1012--1014, 2009.

\bibitem{asur:pred}
S.~Asur and B.~A. Huberman, ``Predicting the future with social media,''
  arXiv.org e-print archive: 1003.5699.

\bibitem{lesk:meme}
J.~Leskovec, L.~Backstrom, and J.~Kleinberg, ``Meme-tracking and the dynamics
  of the news cycle,'' in \emph{Proc. {ACM} {SIGKDD} Intl. Conf. on Knowledge
  Discovery and Data Mining}, 2009.

\bibitem{watt:infl}
D.~J. Watts and P.~S. Dodds, ``Influentials, networks, and public opinion
  formation,'' \emph{Journal of Consumer Research}, vol.~34, no.~4, pp.
  441--458, 2007.

\bibitem{Cha:2010p2742}
M.~Cha, H.~Haddadi, F.~Benevenuto, and K.~Gummadi, ``Measuring user influence
  in twitter: The million follower fallacy,'' 2010.

\bibitem{sper:expl}
D.~Sperber, \emph{Explaining Culture: A Naturalistic Approach}.\hskip 1em plus
  0.5em minus 0.4em\relax Oxford: Blackwell Publishers, 1996.

\bibitem{sper:mode}
D.~Sperber and N.~Claidi{\`e}re, ``Why modeling cultural evolution is still
  such a challenge,'' \emph{Biological Theory}, vol.~1, no.~1, pp. 20--22,
  2006.

\bibitem{gruh:info}
D.~Gruhl, R.~Guha, D.~Liben-Nowell, and A.~Tomkins, ``Information diffusion
  through blogspace,'' in \emph{{WWW2004}: Proc. 13th Intl Conf on World Wide
  Web}, NYC, NY, USA, May 17-22 2004, pp. 491--501.

\bibitem{Balog:2006p2268}
K.~Balog, G.~Mishne, and M.~de~Rijke, ``Why are they excited? identifying and
  explaining spikes in blog mood levels,'' \emph{Proc. 11th Meeting Eur.
  chapter of the Association for Comp. Ling. EACL}, pp. 207--210, 2006.

\bibitem{lloy:news}
L.~Lloyd, P.~Kaulgud, and S.~Skiena, ``Newspapers vs. blogs: Who gets the
  scoop?'' in \emph{AAAI Spring Symposium on Computational Approaches to
  Analyzing Weblogs(AAAI-CAAW), Palo Alto, California, USA}, 2006.

\bibitem{Mishne:2006p2282}
G.~Mishne and M.~de~Rijke, ``Capturing global mood levels using blog posts,''
  \emph{AAAI 2006 Spring Symp. on Computational Approaches to Analysing
  Weblogs}, 2006.

\bibitem{kuma:burs}
R.~Kumar, J.~Novak, P.~Raghavan, and A.~Tomkins, ``On the bursty evolution of
  blogspace,'' \emph{World Wide Web}, vol.~8, pp. 159--178, 2005.

\bibitem{Leskovec:sdm}
J.~Leskovec, M.~McGlohon, C.~Faloutsos, N.~Glance, and M.~Hurst, ``Cascading
  behavior in large blog graphs,'' in \emph{Proc. 7th {SIAM} Intl. Conf. on
  Data Mining (SDM)}, 2007.

\bibitem{coin:soci}
J.-P. Cointet and C.~Roth, ``Socio-semantic dynamics in a blog network,'' in
  \emph{{IEEE} Intl. Conf. Social Computing}, 2009, pp. 114--121.

\bibitem{Java:2006p1951}
A.~Java, P.~Kolari, T.~Finin, and T.~Oates, ``Modeling the spread of influence
  on the blogosphere,'' \emph{Proceedings of the 15th International World Wide
  Web}, p.~7, May 2006.

\bibitem{koss:stru}
G.~Kossinets, J.~Kleinberg, and D.~J. Watts, ``The structure of information
  pathways in a social communication network,'' in \emph{Proc. SIGKDD 08},
  2008.

\bibitem{vale:soci}
T.~Valente, ``Social network thresholds in the diffusion of innovations,''
  \emph{Social Networks}, vol.~18, pp. 69--89, 1996.

\bibitem{zhou:topi}
D.~Zhou, X.~Ji, H.~Zha, and C.~L. Giles, ``Topic evolution and social
  interactions: How authors effect research,'' in \emph{Proc. CIKM'06}, 2006.

\bibitem{adar:impl}
E.~Adar, L.~Zhang, L.~A. Adamic, and R.~M. Lukose, ``Implicit structure and the
  dynamics of blogspace,'' in \emph{Workshop Weblogging Ecosystem, 13th WWW},
  2004.

\bibitem{Java:2006p1948}
A.~Java, ``Tracking influence and opinions in social media,'' \emph{ebiquity},
  p.~69, Nov 2006.

\bibitem{Schmid94}
H.~Schmid, ``Probabilistic part-of-speech tagging using decision trees,'' 1994.

\bibitem{Page1998}
\BIBentryALTinterwordspacing
L.~Page, S.~Brin, R.~Motwani, and T.~Winograd, ``The pagerank citation ranking:
  Bringing order to the web.'' Stanford InfoLab, Technical Report 1999-66,
  November 1999. [Online]. Available:
  \url{http://ilpubs.stanford.edu:8090/422/}
\BIBentrySTDinterwordspacing

\end{thebibliography}
